%% file: MMI.tex
\documentclass[superscriptaddress,prx,reprint,amsmath,twocolumn,amssymb
]{revtex4-1}

\input{MMI_Preamble}

\begin{document}

\title{Multimode interferometry for entangling atoms in quantum networks}

\author{Thomas D. Barrett}
\affiliation{University of Oxford, Clarendon Laboratory, Parks Road, Oxford  OX1 3PU, UK}

\author{Allison Rubenok}
\affiliation{Quantum Engineering Technology Labs, H. H. Wills Physics Laboratory and Department of Electrical and Electronic Engineering, University of Bristol, BS8 1FD, UK}

\author{Dustin Stuart}
\affiliation{University of Oxford, Clarendon Laboratory, Parks Road, Oxford  OX1 3PU, UK}

\author{Oliver Barter}
\affiliation{University of Oxford, Clarendon Laboratory, Parks Road, Oxford  OX1 3PU, UK}

\author{Annemarie Holleczek}
\altaffiliation[now at: ]{Bosch Car Multimedia Portugal S.A., Rua Max Grundig 35, 4705-820 Braga, Portugal}
\affiliation{University of Oxford, Clarendon Laboratory, Parks Road, Oxford  OX1 3PU, UK}

\author{Jerome Dilley}
\affiliation{University of Oxford, Clarendon Laboratory, Parks Road, Oxford  OX1 3PU, UK}

\author{Peter B.\,R. Nisbet-Jones}
\affiliation{University of Oxford, Clarendon Laboratory, Parks Road, Oxford  OX1 3PU, UK}

\author{Konstantinos Poulios}
\altaffiliation[now at: ]{School of Physics \&{} Astronomy, University of Nottingham, University Park, Nottingham NG7~2RD, UK}
\affiliation{Quantum Engineering Technology Labs, H. H. Wills Physics Laboratory and Department of Electrical and Electronic Engineering, University of Bristol, BS8 1FD, UK}

\author{Graham D. Marshall}
\affiliation{Quantum Engineering Technology Labs, H. H. Wills Physics Laboratory and Department of Electrical and Electronic Engineering, University of Bristol, BS8 1FD, UK}

\author{Jeremy L. O'Brien}
\affiliation{Quantum Engineering Technology Labs, H. H. Wills Physics Laboratory and Department of Electrical and Electronic Engineering, University of Bristol, BS8 1FD, UK}

\author{Alberto Politi}
\altaffiliation[now at: ]{School of Physics and Astronomy, University of Southampton, Southampton, SO17~     1BJ, UK}
\affiliation{Quantum Engineering Technology Labs, H. H. Wills Physics Laboratory and Department of Electrical and Electronic Engineering, University of Bristol, BS8 1FD, UK}

\author{Jonathan C.\,F. Matthews}
\affiliation{Quantum Engineering Technology Labs, H. H. Wills Physics Laboratory and Department of Electrical and Electronic Engineering, University of Bristol, BS8 1FD, UK}

\author{Axel Kuhn}
\email{axel.kuhn@physics.ox.ac.uk}
\affiliation{University of Oxford, Clarendon Laboratory, Parks Road, Oxford  OX1 3PU, UK}

\date{\today}

\input{MMI_Abstract}

\maketitle


\input{MMI_Introduction} 
\input{MMI_Source} 
\input{MMI_Chip} 
\input{MMI_Results} 
\input{MMI_Conclusions} 
\input{MMI_Acknowledgements}

\appendix
\crefalias{section}{appendix}
\input{MMI_Appendix_Source}
\input{MMI_Appendix_Similarities}

\bibliographystyle{apsrev4-1}
\bibliography{MMI_References}

\end{document}

%% file: MMI_Preamble.tex
\usepackage{graphicx} 
\usepackage{amsmath} 
\usepackage{xifthen}
\usepackage{physics}

\usepackage{printlen}
\usepackage{todonotes}
\usepackage[normalem]{ulem}

\usepackage{hyperref} 
\usepackage[noabbrev]{cleveref} 
\crefname{appendix}{appendix}{appendices}
\Crefname{appendix}{Appendix}{Appendices}

\usepackage{etoolbox} 
\makeatletter
\appto{\appendix}{%
  \@ifstar{\def\theequation@prefix{A.}}%
          {}%
}
\makeatother

\usepackage[load=prefixed,load=abbr,separate-uncertainty=true]{siunitx} 
\DeclareSIUnit\facet{facet}

\usepackage{xcolor}
\definecolor{mmaGreen}{rgb}{0.560181, 0.691569, 0.194885}
\colorlet{darkgreen}{green!40!black}
\colorlet{darkpink}{pink!80!black}
\colorlet{lightblue}{blue!60!white}

\newcommand{\Rb}{$^{87}$Rb}
\newcommand{\Fin}{\mathcal{F}}
\newcommand{\sub}[1]{_{\mathrm{#1}}}
\newcommand{\DTwo}{$\mathrm{D}_2$}
\newcommand{\pmstack}[2]{\substack{+#1 \\ -#2}}
\newcommand{\gcorr}[2]{g^{\left(#1\right)}\left(#2\right)}
\newcommand{\e}{\mathrm{e}}
\newcommand{\ii}{\mathrm{i}}
\newcommand{\etal}{\emph{et al.}}
\newcommand{\ie}{i.e.}

\newcommand{\NbyN}[2]{$#1$$\times$$#2$}

\newcommand{\conj}[1]{#1^{*}}
\newcommand{\creop}[2]{
	\ifthenelse{\equal{#2}{}}
		{\hat{#1}^{\dag}}
		{\hat{#1}^{\dag}_#2}
	}
\newcommand{\annop}[2]{
	\ifthenelse{\equal{#2}{}}
		{\hat{#1}^{\mathstrut}}
		{\hat{#1}^{\mathstrut}_#2}
	}
\newcommand{\ketvac}{\ket{0}}
\newcommand{\bravac}{\bra{0}}

%% file: MMI_Abstract.tex
\begin{abstract}
We bring together a cavity-enhanced light-matter interface with a multimode interferometer (MMI) integrated onto a photonic chip and demonstrate the potential of such hybrid systems to tailor distributed entanglement in a quantum network.  The MMI is operated with pairs of narrowband photons produced \emph{a priori} deterministically from a single \Rb{} atom strongly coupled to a high-finesse optical cavity.  Non-classical coincidences between photon detection events show no loss of coherence when interfering pairs of these photons through the MMI in comparison to the two-photon visibility directly measured using Hong-Ou-Mandel interference on a beam splitter.  This demonstrates the ability of integrated multimode circuits to mediate the entanglement of remote stationary nodes in a quantum network interlinked by photonic qubits.
\end{abstract}

%% file: MMI_Introduction.tex
\section{Introduction}

Entanglement is an essential resource for many applications of quantum information processing (QIP). In particular the creation of entanglement between remote nodes of distributed quantum networks is a key goal of the field \cite{raussendorf03,nielsen06, jozsa06}.  However, preparing multipartite entangled states is challenging as bringing together distant nodes is often impractical.  Networks of interlinked stationary (typically single atoms or ions) and flying (photonic) qubits offer a scalable route to bridging these physical distances \cite{reiserer15}, but necessitate a reliable interface between these elements.  A single atom strongly coupled to a single mode of the electric field, where the internal spin-state of the atom is entangled \hl{red}{with} the emitted photon polarisation, is an ideal architecture for realising such a system.  The entanglement of distant atoms can then be achieved by leveraging this atom-photon entanglement.  In its simplest form the transfer of a quantum state between two remote atoms can be realised by the exchange of a single-photon \cite{ritter12}, however measurement-induced entanglement swapping actions \cite{moehring07, olmschenk09} on photons emitted from many atoms offer the opportunity to scale up the approach and to create arbitrary entangled states across many network nodes.  Here we present the essential first step, a hybrid system where an \emph{a priori} non-probabilistic source of polarised single-photons, produced from a single \Rb{} atom strongly coupled to an optical cavity, is used to operate a multimode interferometer (MMI) integrated onto a photonic chip.

Coupled atom-cavity systems, with the high degree of control they provide over the light-matter interface, are a versatile tool for QIP \cite{kuhn02,vasilev10,dilley12b,nisbet13}.  Single-photon emission is one application of this interface and is \emph{a priori} deterministic. Photons are produced in well defined quantum states, which goes beyond the intrinsically probabilistic sources, such as those based on spontaneous parametric down-conversion (SPDC), commonly used for proof-of-principle demonstrations towards linear optical quantum computing (LOQC) \cite{holleczek16,barter16}.  The creation of entangled pairs of photons, emitted sequentially from a single atom \cite{wilk07b}, and of atoms, via a photon emitted by one atom and absorbed by the second \cite{ritter12}, have been used to demonstrate atom-photon entanglement within such systems.

One can consider multiple atom-cavity systems, with distinct atomic states before and after the emission of a single-photon, each providing a different input to a multimode interferometer which implements a unitary link between inputs and outputs.  In a time-resolved setting, where the time from photon emission to detection is much shorter than the photon coherence length, any detection event at an output mode projects the ensemble of input channels into an entangled state.  This measurement-induced entanglement relies on the unitary transformation destroying the `which-path' information of the photon emission and persists until every emitted photon has been detected.

In principle any linear optics unitary operation can be realised by a series of \NbyN{2}{2} directional couplers~\cite{reck94}, but such networks become increasingly complex for the creation of larger entangled states.  This is especially true for genuine multipartite entanglement, where the $N$-partite entanglement cannot be described as the mix of $M$-party entangled states ($M{<}N$) and which requires the simultaneous involvement of all $N$ parties to be realised \cite{kimble09a}.  The high degree of information encoded in these highly-entangled states makes them of particular interest to quantum information protocols \cite{knill01, briegel01b, briegel03, hillery99, lloyd96}. Rather than requiring increasing numbers of pairwise entanglement operations to create larger entangled states \cite{kimble10}, a potentially more resource efficient and versatile tool is an MMI, where every input mode is coupled to every output mode through a single waveguide.  In this way an MMI immediately provides the ability to interact any subset of an $N$-photon input state -- providing a platform to efficiently create multipartite entangled states with a suitably designed unitary operation.  MMIs have already been designed to serve a number of purposes such as demultiplexers \cite{hong07,xiao07}, power splitters \cite{feng08}, optical attenuators \cite{jiang05} and optical switches \cite{nagai02,wang06}.  By demonstrating pairwise entanglement operations using an MMI and cavity-photons we show the ability of these integrated circuits to mediate the entanglement operations required for the generation of distributed multipartite entangled states in a real-world environment.  

We emit single photons in well defined quantum states from a single \Rb{} atom coupled to a high-finesse optical cavity, and deterministically route sequential emissions down paths of differing length such that they are simultaneously passed through an MMI integrated onto a photonic chip.  These ultra-narrow-band photons have a correspondingly long coherence time, giving rise to non-classical coincidences between photon detections that are up to three orders of magnitude further apart in time than the propagation time across the chip.  We emphasise that in this work, where the temporal length of the photon wavepackets ($\SI{300}{\ns}$) far exceeds the timing precision of the detectors (${<}\SI{100}{\ps}$), we extend use of the term `coincidences' beyond the conventional meaning of simultaneous detection events.  Instead, we say any pair of detection events with a temporal separation within a time-window of interest -- which typically is the length of the photon wavepackets but this is explicitly changed depending on the analysis -- are coincident. The quantum interference of two photons passed through the MMI is contrasted to the classical behaviour observed when the input pair are made fully distinguishable, and the long temporal length of these photons allows the performance of the chip to be characterised in a time-resolved manner.  As the first detection projects the input modes into an entangled state, which is then subsequently measured by the second detection, the degree to which these coincident detections display the expected behaviour for indistinguishable photons is then a measure of the success with which we prepare and preserve entanglement.  Equivalent performance is observed when operating the MMI to the two-photon visibility our source demonstrates in a simple Hong-Ou-Mandel (HOM) experiment. From this we can conclude that our hybrid system of a cavity-based atom-photon interface and an integrated MMI is suitable for use in distributed quantum networks.

%% file: MMI_Source.tex
\section{Experimental Overview}

\subsection{Single-photon generation}

\begin{figure*}[t!]
\includegraphics{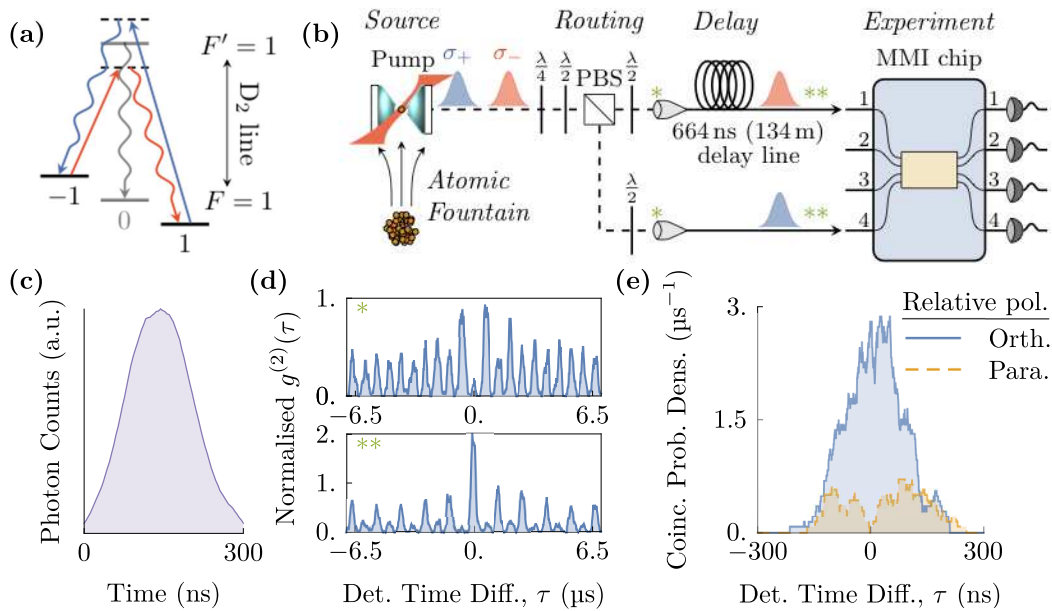}
\caption{(a) Energy level diagram of V-STIRAP processes between magnetic sublevels of the \Rb{} \DTwo{} line.  (b) Experimental set-up of the hybrid source-chip system showing the routing of polarised photons into paths of different length such that pairs of photons are simultaneously delivered to the MMI chip.  (c) Sliding histogram (bin width and pitch of \SI{40}{\ns} and \SI{4}{\ns} respectively) of the count rate for the \SI{300}{\ns} photons emitted from the source. (d) $\gcorr{2}{\Delta\tau}$ measured after \hl{red}{a} Hanbury Brown-Twiss set-up both before the delay lines and with random routing of photons ($\color{mmaGreen}*$) and after the delay lines with deterministically routed photons $(\color{mmaGreen}**$) (the plots use a bin width of \SI{100}{\ns} and a pitch of \SI{20}{ns}). (e) Hong-Ou-Mandel interference of photon pairs on a 50:50 beam splitter, in place of the MMI chip, for parallel (indistinguishable) and orthogonally (distinguishable) polarised photon pairs, shown as a sliding histogram (bin width and pitch of \SI{40}{\ns} and \SI{4}{\ns} respectively).}
\label{fig:source}
\end{figure*}

The polarised single-photons are produced by a V-STIRAP process between magnetic sublevels of the \Rb{} \DTwo{} line, a scheme first presented and demonstrated by Wilk \etal{} \cite{wilk07c,wilk07}.  An external magnetic field lifts the degeneracy of the magnetic sublevels such that the cavity can selectively couple transitions between specific spin-states of the atom, allowing a suitably tuned pump laser to drive Raman transitions from $\ket{F{=}1,m_F{=}{\pm}1}$ to $\ket{F{=}1,m_F{=}{\mp}1}$, resulting in the emission of a $\sigma{^\pm}$ photon into the cavity.  This scheme is illustrated in \cref{fig:source}(a) where it can be seen that the application of pump pulses alternately detuned from the cavity resonance by the splitting of the ground level stretched states results in the emission of a stream of alternately polarised single photons.  Details of the experimental realisation of this driving scheme in our system can be found in \cref{app:singlePhotonSource}.

The second-order correlation function, $\gcorr{2}{\Delta\tau}$, marked as ($\color{mmaGreen}*$) in \cref{fig:source}(d), was measured using a standard Hanbury Brown-Twiss configuration \cite{twiss56,twiss56b} (\ie{} by recording coincident photon detections between detectors at ($\color{mmaGreen}*$) in \cref{fig:source}(b)).  This shows a small, but non-negligible, $\gcorr{2}{0}=0.067$, which is attributable to non-Raman-resonant processes that result in the same atom emitting two photons within one single driving interval.  The correlation rate peaks at times corresponding to detection events an integer number of driving intervals apart.  The largest peaks are measured for sequential emissions and the non-zero possibility that a spontaneous emission during photon production leaves the atom in a `dark' state~\cite{barrett18b}, from which it can produce no more photons, results in reduced correlation rates at longer detection time differences.

The experimental set-up required to operate the MMI chip is shown in \cref{fig:source}(b), where the alternatively polarised photons are routed into paths of differing length.  A \SI{134}{m} fibre spool delays one of the photons by the \SI{664}{ns} duty cycle of the production scheme such that pairs of sequentially emitted photons are delivered simultaneously to the MMI.  Repeating a $\gcorr{2}{\Delta\tau}$ measurement on the coincident detections between these two paths, shown as plot ($\color{mmaGreen}**$) in \cref{fig:source}(d) and measured at ($\color{mmaGreen}**$) in \cref{fig:source}(b), illustrates this routing with photon pairs only arriving in every second driving interval.  The large central peak of detections in the same time interval shows that the polarised single-photon source allows the delivery of photon pairs to be realised with increased efficiency than is possible for random routing.  The smallest correlation peaks in \cref{fig:source}(d)($\color{mmaGreen}**$), corresponding to detection events an odd number of driving intervals apart, are caused by one photon taking the `wrong' path, which is attributable to experimental imperfections in the polarisation state of the emitted photons (caused by a small birefringence of the cavity mirrors \cite{barrett18c}) and in the alignment of the polarisation-routing optical elements.  These effects only result in a limited routing efficiency and so do not impact the overall two-photon interference.

The quantum interference of these photon pairs is characterised by a Hong-Ou-Mandel experiment \cite{hong87,legero04}.  This is achieved by using a 50:50 beam splitter in place of the MMI in the configuration illustrated in \cref{fig:source}(b), and then rotating the relative photon polarisations at ($\color{mmaGreen}**$) to measure the orthogonally (distinguishable) and parallel (indistinguishable) polarised cases.  The cross-detector coincidences as a function of detection time difference at the output of this beam splitter in these two cases are compared in \cref{fig:source}(e).  For parallel polarised photons the suppression of these coincidences illustrates the `bunching' of indistinguishable pairs as they coalesce and exit into the same output mode.  The two-photon visibility is defined as the reduction in likelihood of measuring cross-detector coincidences for parallel polarised photons compared to the non-interfering orthogonally polarised reference.  Measured over the entire interaction time of the \SI{300}{\ns} long photons the visibility is \SI{70.8\pm4.6}{\percent}, which increases to $\geq\SI{97.8}{\percent}$ when considering only detections within less than $\SI{23}{ns}$ of each other.  This temporal variation in the photon distinguishability is a result of their coherence properties, the theory of which is described in detail in \cite{legero06}.  The behaviour observed in our system indicates the interference of narrowband photons with a $2\pi{\times}\SI{2.15}{\MHz}$ bandwidth.  The high two-photon visibility within the long coherence time allows us to use this source to examine the interference within the MMI.

%% file: MMI_Chip.tex
\subsection{MMI photonic chip}

\par 
The MMI on the photonic chip was fabricated using an optical lithography process to form waveguides of silica doped with germanium and boron on a silicon wafer \cite{politi08}.  The rectangular waveguides have a \NbyN{\SI{3.5}{\um}}{\SI{3.5}{\um}} cross-section and a refractive index contrast of $\Delta n=( n^{2}\sub{core} {-} n^{2}\sub{cladding} ) / (2 n^{2}\sub{core} ) \approx \SI{0.5}{\percent}$ to support the fundamental mode at \SI{780}{nm}.  The four input and four output modes are all coupled to a single multimode waveguide \cite{peruzzo11} where the self-imaging principle states that the input field is then reproduced in single (or multiple) images at periodic intervals along the waveguide length \cite{ulrich75, ulrich75b}.  Coupling into and out of the chip is achieved by arrays of polarisation maintaining fibres glued directly to the chip to remove alignment mechanics.

The total loss through the chip varies from \SIrange{3.3}{5.9}{\dB} across the input modes.  We post-select only the experiments where the input photons are detected at the chip outputs such that the relative transmission of each input mode only affects the efficiency of data acquisition.  The relevant transfer matrix, $M$, describing the operation of the chip is then normalised to neglect losses within the chip.  This matrix was directly characterised with coherent light using the approach described in \cite{rahimiKeshari11,rahimiKeshari13}.  Direct transmission measurements give the amplitude elements of $M$, with the relative offset between interference fringes measured at each output mode when driving an interferometer between pairwise combinations of input modes giving the phase information.  The transfer matrix for our MMI was measured to be
\begin{equation}
 M = \pmqty{
 0.28 & 0.7 & 0.45 & 0.48 \\
 0.41 & 0.6\e^{\ii3.67} & 0.41\e^{\ii3.86} & 0.54\e^{\ii1.34} \\
 0.42 & 0.61\e^{\ii2.84} & 0.55             & 0.38\e^{\ii4.29} \\
 0.56 & 0.41\e^{\ii0.39} & 0.59\e^{\ii2.94} & 0.41\e^{\ii4.25}
}.
 \label{eq:mmiM}
\end{equation}

It should be noted that, as we have experimentally measured each matrix element, this is not an ideal unitary matrix.  In principle there exists a larger unitary matrix that accounts for all loss channels and fully describes the action of the MMI as is discussed in detail in \cite{rahimiKeshari13}.  For our purposes however, a well-characterised transfer matrix is sufficient to explain the modified detection statistics of photon pairs passed through the chip.  The transfer matrix is considered to be independent of polarisation of an incident photon in this work, which we will see to be justified in the results presented in \cref{sec:results}.

\subsection{Detection-based quantum state preparation}

Photon detection is performed by superconducting nanowire detectors~\footnote{Photon Spot, model number NW1FC780.} that typically measure with detection efficiencies exceeding \SI{80}{\percent}, a detection jitter of ${<}\SI{100}{\ps}$ and recovery times of ${\sim}\SI{50}{\ns}$.  The measured dark count rates are negligible, ranging from \numrange{5}{66} per hour, and every detection event is recorded at run time with \SI{81}{\ps} precision by a commercial time-to-digital converter~\footnote{Qutools quTAU.} with all data processing deferred to a later time.

\par 
The successive detections of two photons interfered through the MMI sequentially prepares then measures an entangled state of the input modes.  This process can be understood by a step-by-step analysis of the system at each stage.  Two indistinguishable photons input into different modes, $i$ and $j$, prepares the initial state $\ket{\Psi\sub{in}} = \creop{a}{i}\creop{a}{j}\ketvac{}$.  The first detection of a photon in the output mode $k$ -- presuming $\ket{\Psi\sub{in}}$ remains unchanged prior to this detection -- then projects the ensemble of input channels into the state
\begin{equation}
\begin{aligned}
	&{} \annop{b}{k}\ket{\Psi\sub{in}} = \sum_{X} M_{Xk} \annop{a}{k}\creop{a}{i}\creop{a}{j}\ketvac{}, \\
	\xrightarrow{\text{norm}}\ & \frac{M_{ik}\creop{a}{j} + M_{jk}\creop{a}{i} }{\sqrt{\abs{M_{ik}}^{2} + \abs{M_{jk}}^{2} }}\ketvac{} \equiv \ket{\Psi_{\mathrm{ent},k}}, 
	\label{eq:firstPhotonDetection}
\end{aligned}
\end{equation}
where the action of the MMI has been described by the mappings
\begin{equation}
	\creop{b}{i} = \sum_{X} \conj{M}_{Xi}\creop{a}{X}, \quad \annop{b}{i} = \sum_{X} M_{Xi} \annop{a}{X}.
	\label{eq:MMImappings}
\end{equation}
We emphasise that if the coherence length of the photons exceeds the distance between emission and detection, then any entanglement of input modes must be expressed in terms of the internal states of the emitters.  Here, the spin-state of the atom, which flips upon emission, is then entangled with the photon number in the delay line. Accordingly, if multiple atom-cavity systems were used to prepare the input vector, the first photon detection would project the ensemble of emitters into a multi-partite entangled state.  Realising specific forms of distributed entanglement is then a matter of designing an MMI with the appropriate transfer matrix.

For our initial state, $\ket{\Psi\sub{in}}$, the probability of the first detection being in output mode $k$ is given by
\begin{equation}
\begin{aligned}
	P_k &= \frac{ \bra{\Psi\sub{in}} \creop{b}{k} \annop{b}{k}\ket{\Psi\sub{in} } }{ \sum_{X} \bra{\Psi\sub{in}} \creop{b}{X} \annop{b}{X}\ket{\Psi\sub{in} } }
	= \frac{ \abs{M_{ik}}^2 + \abs{M_{jk}}^2 }{ \sum_X ( \abs{M_{iX}}^2 + \abs{M_{jX}}^2 ) }, \\
	&= \frac{1}{2} ( \abs{M_{ik}}^2+\abs{M_{jk}}^2 ).
	\label{eq:probK}
\end{aligned}
\end{equation}
A second photon detection, in channel $l$, then reduces the input and output modes to the vacuum as all photons have been detected,
\begin{equation}
	\annop{b}{l}\ket{\Psi_{\mathrm{ent,}k}} = \sum_{X} M_{Xl} \annop{a}{X} \frac{M_{ik}\creop{a}{j} + M_{jk}\creop{a}{i} }{\sqrt{\abs{M_{ik}}^{2} + \abs{M_{jk}}^{2} }}\ketvac{} \xrightarrow{\text{norm}} \ketvac{}.
	\label{eq:secondPhotonDetection}
\end{equation}
This second detection in $l$, conditioned on a previous detection in $k$, occurs with probability
\begin{equation}
	P_{l|k} = \abs*{ \bravac{}\annop{b}{l}\ket{\Psi_{\mathrm{ent,}k}}}^2 = \frac{ \abs{M_{ik}M_{jl} + M_{jk}M_{il}}^2 }{ 2 P_k }.
	\label{eq:probLcondK}
\end{equation}
Repeating this analysis with the first detection in channel $l$ followed by a second detection in $k$, allows us to define the probability of coincident detections between outputs $k$ and $l$ to be
\begin{equation}
\begin{aligned}
	Q_{ij}^{kl} &= \frac{1}{1+\delta_{kl}} ( P_k P_{l|k} + P_l P_{k|l} ), \\
	&= \frac{1}{1+\delta_{kl}}\abs{M_{ik}M_{jl} + M_{il}M_{jk}}^{2},
	\label{eq:mmiPara}
\end{aligned}
\end{equation}
where $\delta_{kl}$ is the Kronecker delta function required to prevent double counting of the detection orders when $k=l$.
For fully distinguishable photons this becomes \cite{mattle95,poulios14,rahimiKeshari13}
\begin{equation}
	C_{ij}^{kl} = \frac{1}{1+\delta_{kl}}\left(\abs{M_{ik}M_{jl}}^{2} + \abs{M_{il}M_{jk}}^{2}\right).
	\label{eq:mmiPerp}
\end{equation}

%% file: MMI_Results.tex
\section{Results}
\label{sec:results}

\begin{figure}[t!]
\includegraphics{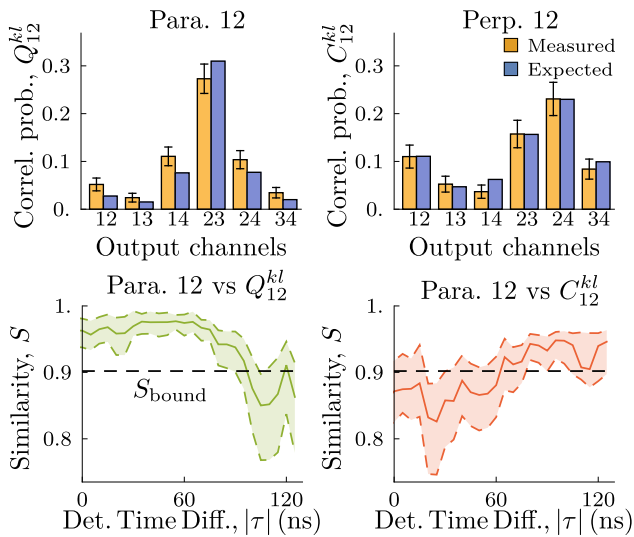}
\caption{Cross-detector coincident detections with photon pairs input into modes 1 and 2 of the MMI.  The upper traces of parallel (Para.~12) and perpendicularly (Perp.~12) polarised input photons show similarities to the expected behaviour of $S=98.9\protect\pmstack{0.4}{0.6}\si{\percent}$ and $S=99.4\protect\pmstack{0.4}{0.5}\si{\percent}$ respectively.  The lower traces show the time-resolved performance of the parallel polarised case by comparing the measured coincidence distributions to the expected behaviour for both interfering (vs $Q_{12}^{kl}$) and non-interfering (vs $C_{12}^{kl}$) photons.  These consider only coincidences within $\abs{\tau}\pm\SI{25}{\ns}$ with the dashed lines marking the \SI{68}{\percent} credible interval.}
\label{fig:resultsInputs12}
\end{figure}

The measured distributions of coincident cross-detector detections for both parallel and perpendicularly polarised photon pairs input into modes 1 and 2 are shown in the upper plots of \cref{fig:resultsInputs12}.  These show similarities to the theoretically predicted distributions of $S=98.9\protect\pmstack{0.4}{0.6}\si{\percent}$ and $S=99.4\protect\pmstack{0.4}{0.5}\si{\percent}$ respectively.  We follow the example of previous work \cite{politi08,holleczek16} by defining
\begin{equation}
	S = \frac{ \sum_i \sqrt{p_i q_i} }{ \sqrt{\sum_i p_i \sum_i q_i} }
	\label{eq:similarity}
\end{equation}
 where $p_i$ and $q_i$ are elements of the compared distributions.  This is the classical fidelity normalised for distributions that do not sum to 1, as is the case here since same-detector coincidences are still neglected.

Whilst identical and orthogonal distributions provide similarities of $S=1$ and $S=0$ respectively, pairs of randomly selected distributions do not produce similarities evenly sampled from between these bounds.  For instance, the similarity between the predicted distributions for indistinguishable and fully distinguishable photon pairs in \cref{fig:resultsInputs12} is $S\sub{bound}=\SI{90.1}{\percent}$, and two random distributions selected from within a six-dimensional parameter space will most likely show a similarity of $87.6\protect\pmstack{7.3\phantom{0}}{11.6}\si{\percent}$ to each other.  Although the single nature of our cavity-photons has already been demonstrated (see \cref{fig:source}(d)), we also cannot surpass this classical bound if we instead considered our input modes to be coherent states with a sufficiently low mean photon number that our detectors could still resolve coincident detection events.  Whilst pairs of single photons input across these modes have no defined phase relation, simulating the behaviour of coherent states requires the consideration of all possible relative phases between the two inputs, from 0 to $2\pi$, which results in any interference effects being averaged out.  As such we would expect the measured behaviour to be exactly that derived for pairs of distinguishable single photons.  That input pairs of parallel polarised single photons exhibit a similarity to the predicted behaviour of indistinguishable two-photon states far exceeding the expectation for classical behaviour, $S\sub{bound}$, verifies that we successfully entangle the two input modes~\cite{terhal00}.

\Cref{app:Similarities} details how the similarities and associated uncertainties are obtained from the measured data.  Further context for the quoted values is also provided and it is of particular note that the chance of a randomly generated distribution exhibiting a similarity to the theoretical predictions within or exceeding the credible interval of the experimentally measured behaviour is \SI{0.40}{\percent} and \SI{0.47}{\percent} for indistinguishable and fully distinguishable photon pairs, respectively.  It would require different random distributions to match our experimentally observed results for every input pairing and relative photon polarisation presented in this work and, with the low probability of this occurring for even one experimental configuration, this clearly shows that quantum interference dominates and explains the photon propagation in the MMI.

The long coherence length of our photons allows us to examine the performance of our hybrid source-chip system in a time-resolved manner.  The lower plots of \cref{fig:resultsInputs12} shows the similarity of the measured \hl{red}{coincidence} distribution for parallel polarised photon pairs to the predicted behaviour for both interfering (vs $Q_{12}^{kl}$) and non-interfering (vs $C_{12}^{kl}$) photons, as a function of the detection time difference.  The behaviour is most similar to the interfering case, as expected, for detections up to \SI{90}{\ns} apart, after which the behaviour moves towards that predicted for non-interfering photons.  This temporal decoherence is a property of the photons and is in agreement with that observed in their HOM interference (\cref{fig:source}(e)).  Photons detected \SI{90}{\ns} apart would correspond to a spatial separation of approximately \SI{18}{\m} when treating them as point particles \footnote{Using the reasonable approximation that the speed of light in a fibre is $2c/3$.}, approximately 1000 longer than the chip itself, showing the two-photon state remains coherent even when the photons can naively be thought to have never been simultaneously present in the chip.  Practically, this is illustrating the lifetime of the entangled state created upon the first photon detection, with no significant decoherence affecting the state within this ${\sim}\SI{90}{\ns}$ window.  It is noteworthy that the time-resolved similarity of the measured data to the expected behaviour for interfering photons never exceeds the equivalent similarity of $S=98.9\protect\pmstack{0.4}{0.6}\si{\percent}$ when considering all coincident detections within the entire span of the photon wavepackets.  This is because smaller data sets are also more susceptible to statistical noise, which is more likely to negatively impact the similarity than positively.


\begin{figure}[t!]
\includegraphics{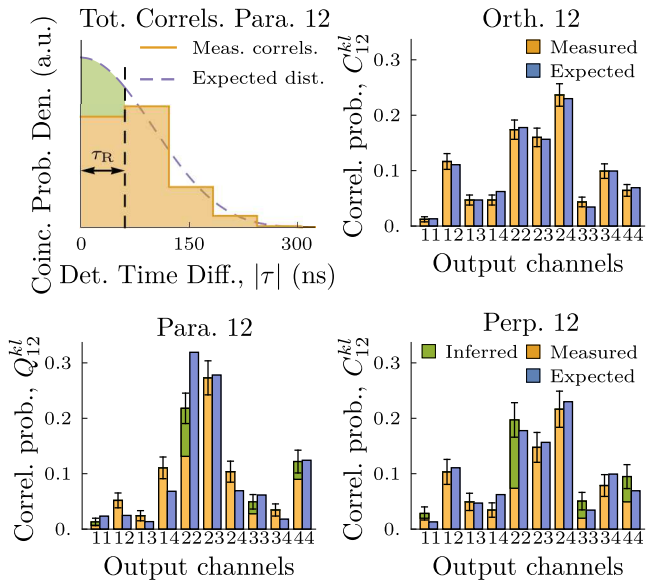}
\caption{Performance of the chip for photon pairs input into modes 1 and 2, corrected for missing same-detector coincident detections due to the finite recovery time, $\tau\sub{R}$.  The upper left plot compares the distribution of all measured coincidences to that expected -- found as described in the text -- from which the total number of missed detections can be inferred.  The distribution of these is known to be the same regardless of the photon interference, and so can be found by looking at the same-detector coincidences observed for orthogonal photons (Orth.~12 with $S=99.6\protect\pmstack{0.2}{0.2}\si{\percent}$) -- `orthogonal' here referring to the property of the photon states not overlapping in time.  Their polarisation is then not relevant, and indeed is parallel for the data shown here.  The lower traces show the complete data, including this same-detector correction, for both parallel (Para.~12, $S=98.3\protect\pmstack{0.5}{0.6}\si{\percent}$) and perpendicularly (Perp.~12, $S=98.9\protect\pmstack{0.5}{0.5}\si{\percent}$) polarised input photons.}
\label{fig:resultsCorrectingCorrs}
\end{figure}

To fully characterise the performance of the MMI, same-detector coincidences, many of which are missed due to the finite recovery time of the detectors, must also be considered.  The distribution of all coincidences, both same- and cross-detector, is given by the autoconvolution of the intensity profile of the photons (seen in \cref{fig:source}(c)).  By matching this profile to the measured distribution of coincidences separated by more than the maximum detector recovery time, $\tau\sub{R}$, we can infer the number of missed coincidences as can be seen in the upper left plot of \cref{fig:resultsCorrectingCorrs}.

To find how these missed same-detector coincidences should be distributed across the output channels, we note from  \cref{eq:mmiPara,eq:mmiPerp} that $Q_{ij}^{kk} / C_{ij}^{kk} = 2$.  As this relationship holds for any choice of input and output modes, the relative distribution of these same-detector coincidences is unchanged by the photon distinguishability.  Thus for a given data set we can find this relative distribution by considering coincidences between photons made fully distinguishable in time.  The upper right plot of \cref{fig:resultsCorrectingCorrs} shows these when considering detection time differences of twice the duty cycle of the experiment, corresponding to the nearest side-peaks to the central peak for the $\gcorr{2}{\Delta\tau}$ measured with photon routing and delay lines in place in \cref{fig:source}(d).  Physically these are events where the production of the second photon succeeds on the second attempt at producing that polarisation and so effectively there are two duty cycles between detections.  As this is much longer than the detector recovery times, no same-detector coincidences are missed in this case.

The lower traces in \cref{fig:resultsCorrectingCorrs} show the complete corrected performance of the MMI for both parallel and perpendicularly polarised photons input into modes 1 and 2 measured across all possible output pairings.  These have similarities (or equivalently fidelities as our distributions now sum to one) to the theoretical predictions of $S=98.3\pmstack{0.5}{0.6}\si{\percent}$ and $S=98.9\pmstack{0.5}{0.5}\si{\percent}$ respectively.  That the behaviour for non-interfering photons is equivalent, within error bars, for both parallel polarised photons distinct in time and perpendicularly polarised photons simultaneously incident on the chip justifies our previous assertion that that there is not a measurable polarisation dependence of the transfer matrix.  It should also be noted that although $Q_{ij}^{kk} / C_{ij}^{kk}$ is a constant value (and this is sufficient to allow us to infer the distribution of missed same-detector coincidences), it deviates from $Q_{ij}^{kk} / C_{ij}^{kk}=2$ in the expected data presented.  This is a result of a small renormalisation \footnote{Any non-unitary transfer matrix predicts probabilities of coincident detections that do not sum to unity across all possible pairs of output modes, necessitating a renormalisation.  In our case the average magnitude of this correction is \SI{1.9}{\percent}.} of the coincidence distribution predicted for interfering photons, required due to the experimentally-measured transfer matrix, $M$, being non-unitary.


\begin{figure}[t]
\includegraphics{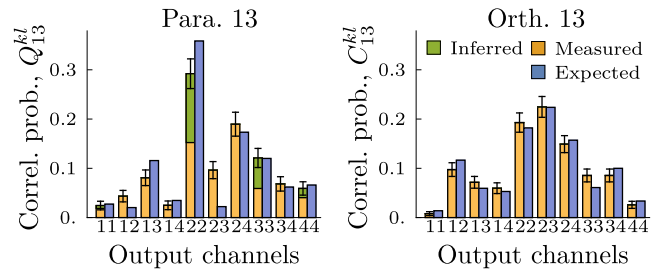}
\caption{Performance of the chip measured for photon pairs input into modes 1 and 3, including a correction for the inferred missed same-detector coincidences.  The distribution of coincidences separated by less than the photon length for parallel polarised input photons (Para.~13) shows a similarity to the expected behaviour of $S=97.7\protect\pmstack{0.6}{0.6}\si{\percent}$.  Orthogonal input photons made distinguishable in time (Orth.~13) taken from the same experimental run show $S=99.4\protect\pmstack{0.3}{0.4}\si{\percent}$.}
\label{fig:resultsAllCorrsInputs13}
\end{figure}

The chips operation with photons input to modes 1 and 3 is summarised in \cref{fig:resultsAllCorrsInputs13} where the parallel polarised photons show $S=97.7\pmstack{0.6}{0.6}\si{\percent}$ to the predicted behaviour.  Orthogonal input photons, made distinguishable in time, show $S=99.4\pmstack{0.3}{0.4}\si{\percent}$.


The performance of our hybrid chip-source system clearly indicates the successful operation of the MMI with cavity-photons.  To identify the origin of the rather small imperfections observed, we note that the interference will always be inherently limited by the coherence properties of the photon pairs, which was directly measured in the HOM presented in \cref{fig:source}(e).  For an input pair with two-photon visibility, $V$, we can consider that the expected MMI output is simply the weighted average of the limiting cases,
\begin{equation}
	R_{ij}^{kl}\left(V\right) = V \cdot Q_{ij}^{kl} + (1-V)\cdot C_{ij}^{kl},
	\label{eq:mmiDistinguishability}
\end{equation}
where $0 \leq V \leq 1$, with $V=0$ ($V=1$) corresponding to completely distinguishable (indistinguishable) photons.
	
The distributions $R_{12}^{kl}\left(V_{12}\right)$ and $R_{13}^{kl}\left(V_{13}\right)$ show maximum similarities to the observed behaviour of parallel polarised photons input to modes 1 and 2 (see \cref{fig:resultsCorrectingCorrs}) and modes 1 and 3 (see \cref{fig:resultsAllCorrsInputs13}) of $S=99.1\pmstack{0.3}{0.4}\si{\percent}$ and $S=99.4\pmstack{0.2}{0.3}\si{\percent}$ respectively for $V_{12}=0.728$ and $V_{13}=0.684$.  This is in good agreement with the two-photon visibility of \SI{70.8\pm4.6}{\percent} observed from the HOM interference of our photons, from which we can conclude that the performance of our system is not impacted by any additional decoherence arising from the interfacing of the atom-cavity source with the photonic MMI chip.

%% file: MMI_Conclusions.tex
\section{Conclusions}
Observing similarities with the expected quantum behaviour in the MMI output exceeding \SI{98}{\percent} for coincident detections several tens of \si{\ns} apart impressively shows the potential of our hybrid approach. Firstly, this clearly demonstrates the uniquely quantum behaviour in the multimode interferometry of cavity photons. Secondly, the long time span between detections demonstrates that the MMI effectively projects the entire input vector into an entangled state upon the first detection, which survives and preserves its coherence until the second photon is detected. To this respect, the similarity recorded as a function of detection-time difference (see \cref{fig:resultsInputs12}) can be regarded as the fidelity of remaining in the entangled state probabilistically prepared upon the first photon detection. 

The presented study demonstrates the feasibility of preparing tailored multipartite entanglement without increased experimental overhead for more highly entangled states.  In our work this entanglement is between the atom-cavity system, and the photon state stored in an optical fibre. For future quantum networking, we anticipate the implementation of a network of atom-cavity systems, all coupled simultaneously by fibres to a photonic MMI chip. The first detections in the MMI output will then project the ensemble of input atoms into a spatially distributed multipartite entangled state.  To allow entanglement to persist after all photons are detected, the atom-cavity system must entangle the internal spin-state of the atom with the emitted photon polarisation, requiring a polarised scheme such as ours. Different measurement outcomes that depend on the distinguishability, and hence interference through the MMI, of the input multi-photon state can then probabilistically project the atoms into an entangled state.

The scalability of these more complex networking proposals is directly related to the total losses in the system.  With an atom coupled to the cavity our overall efficiency, for photon emission, transmission and detection, is $\eta=\SI{9.4\pm0.2}{\percent}$, which is demonstrably sufficient for a proof-of-principle entanglement scheme using pairs of photons.  However, as the likelihood of observing $N$-photon events scales with $\eta^{N}$, higher photon number experiments would require circumventing the present limitations in the photon generation efficiency~\cite{barrett18b}.   These limitations are well understood and we estimate that new schemes accounting for these effects would allow the efficient delivery of 3- and 4-photon states to the chip.  The on-chip losses can also be improved, with ${<}\SI[per-mode=symbol]{1}{\dB\per\facet}$ consistently observed through similar chips in the lab.  As such, it is realistic to believe the presented proof-of-principle experiment could be scaled up to higher photon number.

In summary the combination of our single-photon source with an integrated photonic chip, each of which was individually characterised, has been shown to operate flawlessly, up to imperfections inherent to each individual component, and in full accordance with expectations.  This is an important step towards a distributed quantum network of cavity-based emitter-photon entanglers as it experimentally proves the ability, in a real-world setting, of integrated MMIs to mediate multipartite entanglement across numerous distant nodes in a quantum network.  This is also the first time an atom-cavity single-photon source that provides polarisation control has been integrated with a more complex optical network than a simple beam-splitter.  This control could be combined with readily available fast-switching electro-optical elements such as Pockels cells to allow active routing of the photons, and thus the non-probabilistic preparation of even larger Fock states~\cite{wang17, pittman02b}.  This is a promising approach to surpassing the random photon routing that has limited the photon number of equivalent experiments with previous cavity-based \cite{holleczek16} and SPDC sources \cite{peruzzo11,poulios13,poulios14}, and towards truly scalable quantum networks.

%% file: MMI_Acknowledgements.tex
\begin{acknowledgments}
We acknowledge support for this work through the quantum technologies programme (NQIT hub).  The authors are also grateful to Chris Sparrow for helpful discussions.
\end{acknowledgments}

%% file: MMI_Appendix_Source.tex
\section{Single-photon source}
\label[appendix]{app:singlePhotonSource}

The experimental set-up is shown in \cref{fig:source}(b).  \Rb{} atoms are loaded into a magneto-optical trap (MOT) ${\sim}\SI{8}{\mm}$ below the cavity for $\SI{500}{\ms}$ where they are cooled to ${\sim}\SI{20}{\micro\kelvin}$.  They are then stochastically loaded by an atomic fountain (further details can be found in \cite{nisbet11}) -- an upwardly launched MOT with sufficiently diffuse density upon reaching the cavity that in general only one or zero atoms are loaded at any given time.  At typical launch velocities of ${\sim}\SI{1}{\m\per\s}$ a single atom takes ${\sim}\SI{60}{\us}$ to transit the  \SI{27}{\um} waist of the cavity mode.  This corresponds more than \num{100} attempts at producing a single photon and as such the atom can be considered to be effectively stationary within the duration of a single driving pulse. As the atomic cloud transits the cavity a sequence of \num{20000} driving pulses, each \SI{300}{\ns} long with a $\sin^{2}$ amplitude profile, attempts to produce a stream of alternately polarised single photons.  The driving laser is injected from the side of the cavity and is linearly polarised orthogonally to the mode, such that it decomposes into an equal superposition of $\sigma^{+}$ and $\sigma^{-}$ in the cavity basis.

The driving pulses are separated by \SI{664}{\ns} corresponding to a photon emission rate of ${\sim}\SI{1.5}{\MHz}$.  However as the atomic flux through the cavity is kept intentionally low -- with typically the order of \num{20} atoms passing through the cavity mode per MOT launch -- the system is only intermittently operating at this rate.  Combined with losses through the network, particularly in photonic chip, this required that each data set presented in this work be acquired over multiple hours of experimental run time (though the cavity only contains an atom, and thus has the potential to emit a photon, for ${\sim}\SI{0.2}{\percent}$ of the this time).  As an example, the data presented in \cref{fig:resultsInputs12,fig:resultsCorrectingCorrs} for indistinguishable photon pairs input into modes 1 and 2 of the MMI (`Para.\ 12') consists of \num{247} coincident detection events recorded over \SI{290}{\min} of experimental run time.  Almost every event recorded corresponds to performing the desired experiment as the low background on the superconducting nanowire detectors, with the measured dark count rate ranging from \numrange{5}{66} per hour across our multiple detectors, gives a correspondingly high signal-to-noise ratio. 

The cavity itself is \SI{339}{\um} long and is comprised of two highly reflective mirrors with $R\sub{cav}=\SI{5}{\cm}$ radii of curvature and differing transmissions at \SI{780}{\nm} of approximately \SIlist{4;40}{ppm} for directional emission of the photons.  The measured finesse of $\Fin{} = \num{117800\pm200}$ corresponds to average scattering losses per mirror of \SI{6.6\pm0.1}{ppm} and the cavity linewidth is $\Delta\omega\sub{FWHM}/2\pi=\SI{3.750\pm0.006}{\MHz}$.  It is essential that the Zeeman splitting, $\pm\Delta\sub{Z}$, of the $m_F{=}{\mp}1$ stretched states from the $m_F{=}0$ sublevel of $\ket{F{=}1}$ ground level exceeds this linewidth in order that the cavity can selectively couple to individual spin-states.  The optimum Zeeman splitting used in this work, $\abs{\Delta\sub{Z}}/2\pi{=}\SI{14}{\MHz}$, was determined experimentally to maximise single-photon production efficiency and indistinguishability.  This can be understood as the compromise between the larger field strengths required to sufficiently split the ground state sublevels and the lower field strengths at which single-photon production is minimally impacted by the nonlinear Zeeman effects discussed in \cite{barrett18b}.

The cavity also has non-negligible birefringence with elliptical polarisation eigenmodes split by $\Delta\sub{P}/2\pi=\SI{3.471\pm0.004}{\MHz}$.  A detailed consideration of how this impacts the photon-production scheme can be found in \cite{barrett18c}, however the effects are limited to lowering the effective atom-cavity coupling and imperfect polarisations of the emitted photons.  For the experiments presented here, where we post-select upon coincident detections to only consider photons emitted with the desired polarisations using a combination of routing optics and delay fibres, these are simply inefficiencies in the preparation of the input state to the MMI and so do not effect the measured results.

  The coupling rates of the system, neglecting cavity birefringence and nonlinear Zeeman effects, are then $\{g,\kappa,\gamma\}/2\pi = \{4.77,1.875,3\}\,\si{\MHz}$, where $g$ is the atom-cavity coupling rate, $\kappa$ is the cavity field decay rate and $\gamma$ is the atomic amplitude decay rate.

%% file: MMI_Appendix_Similarities.tex
\section{Similarities}
\label[appendix]{app:Similarities}

\begin{figure}[h]
\includegraphics{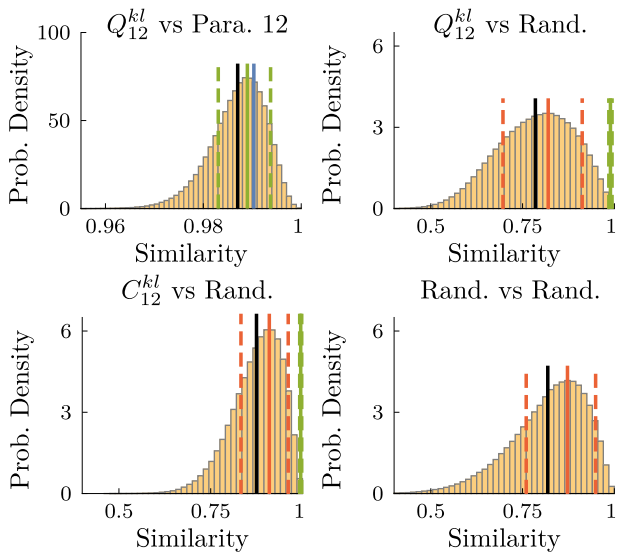}
\caption{Similarities obtained when comparing the theoretical distributions of cross-detector coincident detections for indistinguishable ($Q^{kl}_{12}$) or distinguishable ($C^{kl}_{12}$) photon pairs input into modes 1 and 2 of the MMI to many different distributions.  The upper left plot -- from which we find our quoted similarity and its associated uncertainty as described in the text -- considers the expected behaviour if we were to repeat our measurement, many times with multiple simulated outcomes sampled from the Poissonian statistics that underlie our measured result (Para. 12).  The remaining plots contrast the measured results to truly random behaviour by considering many distributions evenly sampled from a six-dimensional parameter space (Rand.).  Each histogram contains \num{1000000} trials.  `$Q^{kl}_{12}$ vs Para.\ 12' has the mean (black) and most likely (green) similarities obtained marked with vertical lines, along with the \SI{68}{\percent} credible interval (dashed green).  The similarity to the theorey of the single experimentally measured distribution (blue) is also shown.  The remaining plots, comparing theoretical predications to random distributions, are equivalently marked, with the most likely similarity and error bounds marked in red to distinguish them from the corresponding experimentally measured results.}
\label{fig:contextualisingSimilaritiesInputs12}
\end{figure}

On the timescales required to experimentally measure the distribution of coincident detections for a given pair of input modes, the photon detections and associated coincident events exhibit Poissonian counting statistics.  A Poissonian distribution of mean and variance $\lambda$ gives the probability of measuring $n$ events as \cite{hughes10}
\begin{equation}
	P(n;\lambda)=\frac{\e^{-\lambda}\lambda^{n}}{n!}.
	\label{eq:poissonianDistribution}
\end{equation}
If we measure the total number of coincident detections across pairwise combinations of the output modes to be $\{N_1,N_2,\dots\}$, the most likely Poissonian distribution from which each measurement is taken is that with $\lambda=N_i$.  The uncertainty associated with each measured number of events is then $N_i \pm \sqrt{N_i}$ \footnote{The approximation of the $\pm \sigma$ uncertainty on a measured number of events displaying Poissonian statistics as $N\pm\sqrt{N}$ becomes less valid for lower $N$ as the underlying distributions become increasingly asymmetric.  However this is deemed to be sufficient for visually presenting experimental error bars in \cref{fig:resultsInputs12,fig:resultsCorrectingCorrs,fig:resultsAllCorrsInputs13} and a more sophisticated approach is used to propagate these uncertainties through the calculation of the similarities of experiment to theory, as detailed in \cref{app:Similarities}.}, which bounds values within one standard deviation, $\sigma$, of the mean and corresponds to the error bars presented on the experimental data in \cref{fig:resultsInputs12,fig:resultsCorrectingCorrs,fig:resultsAllCorrsInputs13}.

To find the associated uncertainty of the similarity of these measured distributions to the theoretical predictions we use Monte Carlo methods, sampling many possible distributions $\{n_1,n_2,\dots\}$ from $\{P(n_1;N_1),P(n_2;N_2),\dots\}$ and calculating the corresponding distribution of similarities.  From this we can find the  most likely similarity to the theoretical predications and place reasonable bounds upon it.

As an example, the upper-left plot in \cref{fig:contextualisingSimilaritiesInputs12} shows the distribution of similarities obtained when applying this treatment to the measured cross-detector coincidences for indistinguishable photon pairs input into modes 1 and 2 of the MMI (the corresponding experimental data is presented in \cref{fig:resultsInputs12}).  The similarity of the measured distribution of raw events, $\{N_1,N_2,\dots\}$, to the theoretical prediction is marked with the vertical blue line and does not correspond to the most likely similarity obtained from our distribution, which is marked in green.  This is because a normalisation of the raw distribution is inherent to the similarity calculation (recall \cref{eq:similarity}) and this provides a non-trivial relationship between the previously independent values of the number of coincidences across each pair of output channels.  We use the most likely similarity, $S=98.9\protect\pmstack{0.4}{0.6}\si{\percent}$, as our quoted result.  The error bounds, marked by the dashed green lines, correspond the highest posterior density interval containing \SI{68}{\percent} of the distribution -- the approximate certainty contained within the $\pm \sigma$ bounds of a Gaussian (or, for sufficiently large $N$, Poissonian) distribution.  

The upper-right and lower-left plots in  \cref{fig:contextualisingSimilaritiesInputs12} show the similarity of many randomly sampled six-parameter distributions (corresponding to the six possible cross-detector coincidence channels) of the theoretical cross-detector coincidence distributions for indistinguishable and distinguishable photon pairs, respectively, input in modes 1 and 2.  The similarities of pairs of randomly sampled six-parameter distributions is also shown in the lower-right plot.  For interfering photons the most likely similarity to a random distribution is found to be $S=82.1\pmstack{9.3\phantom{0}}{12.3}\si{\percent}$, with the non-interfering case showing $S=90.8\pmstack{5.4}{7.5}\si{\percent}$.  It is unsurprising that values are this high as two randomly chosen distributions most likely have a similarity of $87.6\protect\pmstack{7.3\phantom{0}}{11.6}\si{\percent}$.

Considering how well random distributions match the theoretical performance of our system allows us to conclude with confidence that we are truly measuring the predicted quantum effects.  The chance of a random distribution providing a similarity within or exceeding the \SI{68}{\percent}(\SI{95}{\percent}) credible interval of the experimentally measured performance is \SI{0.40}{\percent}(\SI{0.80}{\percent}) and \SI{0.47}{\percent}(\SI{1.39}{\percent}) for the cases of indistinguishable and distinguishable input pairs, respectively.  This can be seen on plots `$Q^{kl}_{12}\text{ vs Rand.}$' and `$C^{kl}_{12}\text{ vs Rand.}$' where the \SI{68}{\percent} credible intervals of the corresponding experimental results are shown in green.  Furthermore, given the similarity between the two theoretical predictions in this case is $S\sub{bound}=\SI{90.1}{\percent}$, no single random distribution could simultaneously mimic both behaviours with the measured accuracy.